# Stellar Coronal Mass Ejections with HWO

# A Science Case Concept

Authors: R. O. Parke Loyd (astroparke@gmail.com), James Paul Mason (james.mason@jhuapl.edu), Ivey Davis, Kevin France, Meng Jin, Karin Dissauer

Endorsed by: Vladimir S. Airapetian, Christina M.S. Cohen, Yue-Hong Chen, Arika Egan, Alison Farrish, James L. Green, Kosuke Namekata, Yuta Notsu, Antígona Segura, Evgenya Shkolnik, Xudong Sun, Zheng Sun, Astrid M. Veronig, Aline A. Vidotto, Maurice Wilson

## Information for Potential Endorsers

**Deadline** We will stop collecting endorsements and upload a final version of this document to arXiv on 2026 July 15.

**Background** HWO included a formalized process for drafting and aggregating Science Case Development Documents during the summer of 2025. While the period for formal submissions has passed, members of the HWO design community maintain an open ear to scientific developments. One such development is a growing number of ways in which HWO might prove sensitive to stellar coronal mass ejections (CMEs). The authors, and TBD endorsers, feel it is important that the science case for detecting CMEs be represented in the HWO portfolio as mission development continues, particularly given their influence on planetary atmospheres and life. This document constitutes an off-cycle contribution to HWO's science case, to be made available on arXiv for perusal by anyone within or outside of the HWO community.

**Instructions** In the spirit of a white paper, we will accept endorsements of this document, to be listed alphabetically on a final version and added as authors on the arXiv submission. If you wish to endorse, please submit your name as you wish it to be printed on this document via this Google form
https://forms.gle/YByK22Yj47NMEPbp8

# Abstract

The primary mission of the Habitable World Observatory (HWO) will be to constrain the prevalence of life on Earth-like planets. These planets will be subject to impacts by energetic particles generated from coronal mass ejection (CME) shocks that can dramatically deplete ozone, a key biosignature gas. Other biosignatures are also likely vulnerable, though not yet studied. Here, we make a conceptual case for factoring sensitivity to stellar coronal mass ejections into the design of HWO. We drive design considerations by requiring that HWO constrain the rate of CMEs producing ≥ 10% impacts on total ozone column to fewer than one per decade, the timescale over which ozone returns to pre-event levels. As CME detection methods, we consider coronal dimming, doppler shifted emission, high contrast imaging, and planetary aurora as means of detecting CMEs. We explore coronal dimming most thoroughly of the four, though each of these may be possible with HWO with appropriate design considerations.

# Disclaimer

This document follows the template created for the Science Case Development Documents (SCDD) solicited by the HWO development community in the summer of 2025. However, it was developed and posted to arXiv after the submission window for SCDDs closed.

# Main Document

# Step 1: Science Goal

Which stellar systems are the most conducive to the development of life?

*As constraining the occurrence rate of life is HWO's central goal, the importance of which has been elegantly and thoroughly expounded in other documents, we have not attempted to rephrase that motivation here.*

# Step 2: Science Objective

Characterize the intensity of coronal mass ejections (CMEs) produced by the hosts, or analog hosts, of exoplanets targeted for biosignature searches.

## Why Characterizing CMEs is Necessary to Achieve HWO's Life-Finding Mission

Had Earth been subjected to more intense space weather over the course of its life, it might have lost its water-sustaining atmosphere (e.g., Airapetian et al. 2020; Dong et al. 2017; Lammer et al. 2007; France et al. 2020) and life as we know it would not have formed. The Sun, a G2 type star with relatively benign magnetic activity, likely produces less intense space weather than typically more active G-type stars, as well as more active K- and M-type stars that far outnumber G-types.

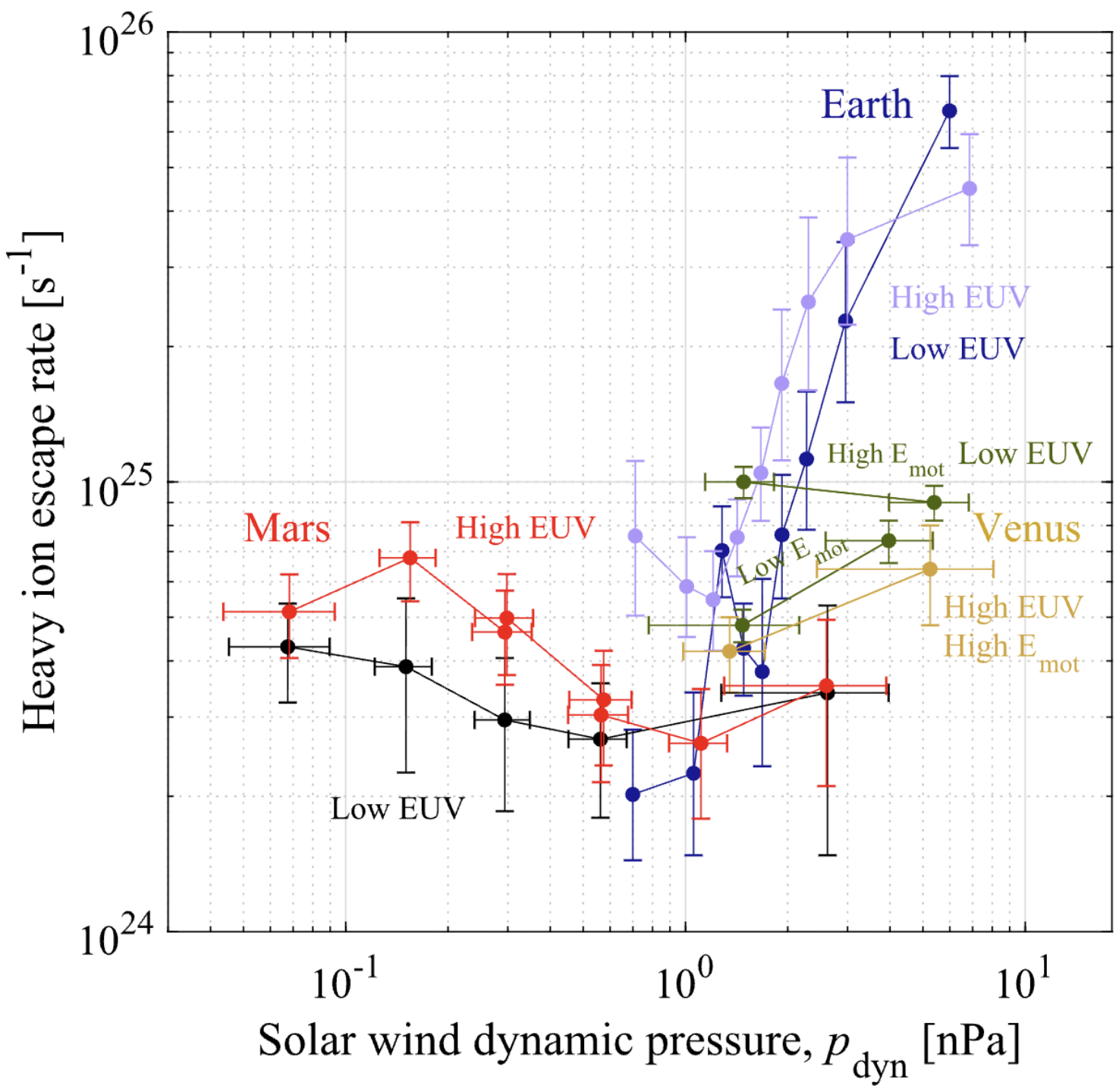

**Figure 1:** Jumps in solar wind dynamic pressure resulting from coronal mass ejections (CMEs) strongly influence heavy ion escape from rocky planets in the solar system (Figure 8 from Ramstad & Barbarash).

As a result of more intense space weather, many planetary systems that may appear excellent candidates for harboring life by metrics such as temperature, size, and age, could, in reality, exist above the "cosmic shoreline," having lost their atmospheres in the face of the space weather they experienced. For example, simulations of the stellar wind of Proxima Centauri and its interaction with the atmosphere of its habitable planet suggest that complete atmospheric stripping is possible, if not likely (Dong et al. 2017). Within the solar system, oxygen loss from Venus, Earth, and Mars has strong dependence on the intensity of the solar wind (Figure 1), especially impulsive spikes due to coronal mass ejections (CMEs; Ramstad & Barabash 2021). The brief and occasional CMEs that strike Mars’ and Venus’ may drive as much as half of their overall mass loss (Edberg et al. 2011, 2010; Luhmann et al. 2007), implying that planets orbiting stars where both the baseline stellar wind and CMEs are more intense may be severely impacted.

The retention of atmospheres will influence which planets HWO searches for signs of life. HWO’s current mission strategy includes an initial step to confirm that candidate target planets retain a substantial atmosphere. This will filter out CME-stripped worlds.

Nonetheless, CMEs remain a critical factor in HWO’s core life-finding mission. Those planets retaining atmospheres targeted for deeper study by HWO will be subject to nonthermal chemistry arising especially from stellar energetic particles (SEPs). SEPs can affect both climate and biosignatures (e.g., Airapetian et al. 2016; Chen et al. 2021, 2025; Khodachenko et al. 2007; Kobayashi et al. 2026; Lammer et al. 2018; Segura et al. 2010; Tabataba-Vakili et al. 2016; Tilley et al. 2019; Venot et al. 2016). Most notably, $NO_x$ species formed from nonthermal particle chemistry can catalyze the destruction of ozone, a key biosignature gas for which HWO is likely to search, over long timescales (Figure 2). SEP effects on other biosignature gasses, like $CH_4$, could be just as significant, but have not been studied as closely as ozone.

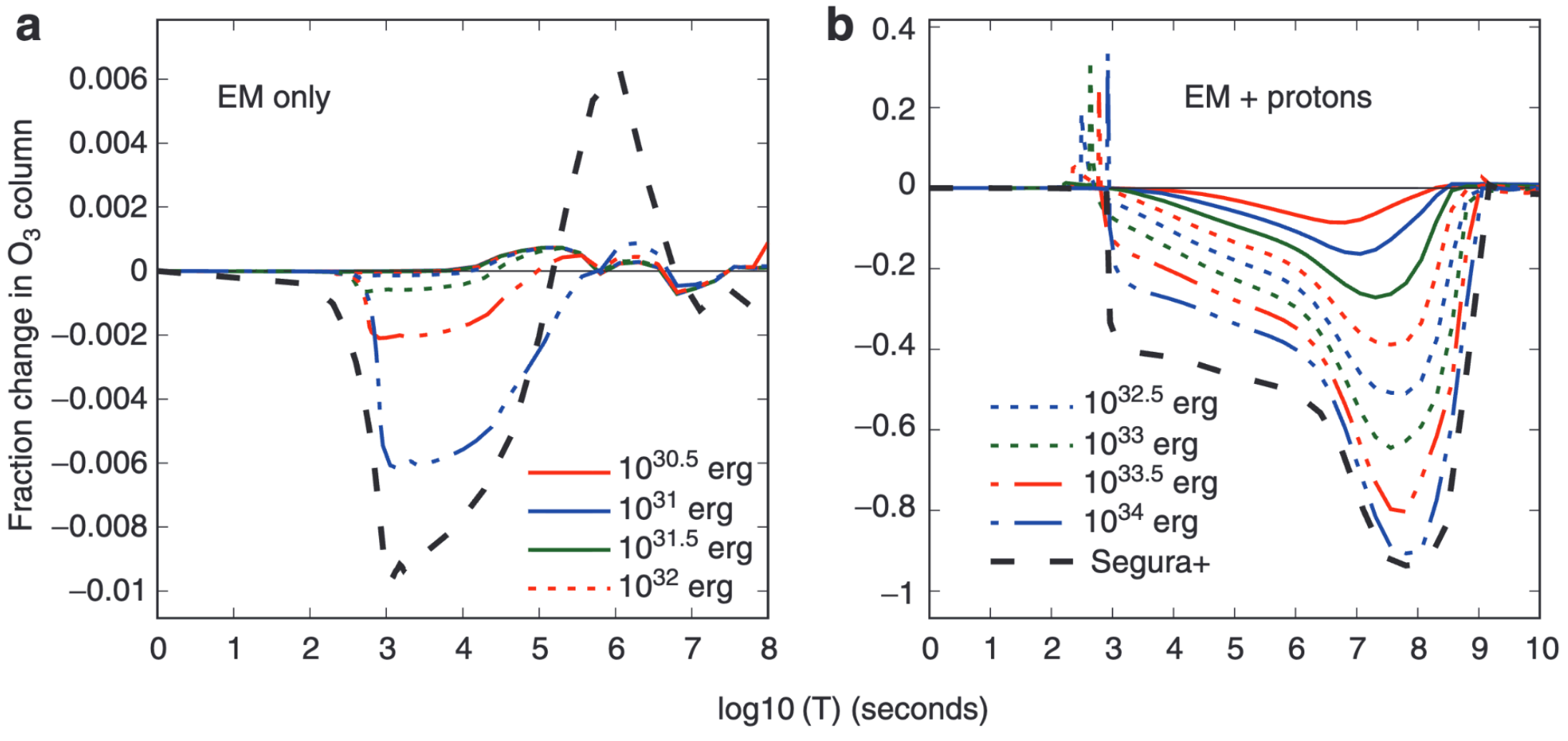

**Figure 2:** Models of the chemical impact of SEPs on an Earth-like planet orbiting an M dwarf star. SEPs were incorporated into a chemical model in an implicit manner by scaling $NO_x$ production rates associated with solar flare/CME/SEP events. The largest solar event observed had an energy of ~$10^{33}$ erg. The panels compare ozone changes in response only to flare electromagnetic radiation (EM; left panel) to radiation and SEPs (EM + protons; right panel). Figure from Tilley et al. 2019.

CMEs generate SEPs through leading shocks that form as CMEs expand at super-Alfvénic speeds. Around half of solar CMEs generate shocks (Chi et al. 2016).

While these persist, they continuously launch SEPs into broad swaths of interplanetary space.

Shock-generated SEPs are thought to be the most impactful to planetary atmospheric chemistry. While SEPs can also be accelerated by magnetic reconnection that takes place during flares and CMEs, these events are generally more tightly collimated, lowering the probability of impacting a planet, and shorter in duration. Their comparatively limited spatial and temporal extent make them a lesser threat than shock-generated SEPs.

If the timescale between major space weather events is sufficiently small, an exoplanet's atmosphere may exist in a constantly fluctuating chemical state. This makes knowledge of a star's CME behavior critical for the interpretation of biosignatures. Hence, HWO's search for life will require constraints on the CME-bombardment a planet experiences to fulfill its core life-finding mission. Indeed, one of the first questions to be asked when chemical imprints of life are discovered will be "what space weather is that planet experiencing?" That question will encompass both a planet's recent history and lifelong cumulative effects.

Unusually weak space weather might also have negative implications for life. Space weather, in particular SEPs and extrinsically-sourced particles that are even higher in energy, galactic cosmic rays (GCRs), have been shown to produce simple precursor biological molecules through chemical models and laboratory experiments (e.g., Airapetian et al. 2016; Kobayashi et al. 2023, 2026; Lingam et al. 2018; Raeside et al. 2025). These also generate complex organic hazes (e.g. Tomasko et al. 2005; Yung et al. 1984) that can determine the detectability of atmospheric spectra (e.g., Espinoza & Perrin 2025), influence planetary climate through albedo (Gao et al. 2021), and provide feedstocks for biochemistry (Moran et al. 2020).

Over the last few decades, constraints on visible space weather, the portion of space weather consisting of X-ray through UV electromagnetic radiation, both quasi-steady and stochastically flaring, have steadily grown. Like stellar winds and CMEs, visible space weather has strong impacts on atmospheric retention and chemistry, and many unknowns still remain. Recognizing this, HWO's current plans include substantial activity monitoring of the hosts of targeted planets to provide context necessary to interpret measured atmospheric compositions.

In comparison, invisible space weather—especially CMEs and the SEP events produced from their shocks—remains one of the least-constrained factors in the genesis, evolution, and detectability of life. Overlooking it risks ignoring a potentially decisive factor in HWO's life-finding mission. Components of HWO's existing design space could be leveraged with emerging techniques to constrain host-star CMEs, likely in tandem with existing plans for stellar activity monitoring. With further development, new approaches may also unfold. Breakthrough space weather science that has both intrinsic value to astrophysics and extrinsic value to the search for life, is possible with HWO.

## Step 3: Physical Parameters

Constrain the rate, velocity, mass, and energy of coronal mass ejections above a threshold that impacts the interpretation of the ozone biosignature.

**Rationale**

Although many solar CMEs and associated SEPs produce observable effects in Earth's atmosphere, including conspicuous aurorae, only the rarest events, approaching energies of ~$10^{33}$ erg, have compositional impacts significant to HWO's biosignature search. Notably, the infamous ~$10^{33}$ erg Carrington event of the 19th century reduced ozone columns as much as 10% in portions of Earth's atmosphere according to a chemical model (Calisto et al. 2013). Indications of more energetic events exist in Earth's geological record and in observations of active stars (Miyake et al. 2012; Papaioannou et al. 2023; Namekata et al. 2025).We adopt $10^{33}$ erg flares as the "threshold of influence" for CMEs bombarding Earth-like planets orbiting at 1 AU.

This $10^{33}$ erg threshold must be adjusted based on variations in the habitable zone distance for varying stellar types to yield an equivalent energy flux experienced by habitable zone planets. For example, for Trappist-1 where the inner edge of the habitable zone is approximately 0.03 AU, the threshold of influence is $10^{30}$ erg. These events might be common for Trappist-1. Recent infrared observations of the star indicate that flares with energies > $10^{30}$ occur with a cadence greater than several times per day (Howard et al. 2023). CMEs, like flares, are products of magnetic reconnection, so Trappist-1's flares may indicate that CMEs with similar energies occur at similar rates.

However, it is not safe to use flares as a proxy for CMEs. The two do not reliably occur in tandem on the Sun, and solar correlations that have been made between flares and CMEs on the Sun yield implausible rates of energy and mass loss for active stars (Drake et al. 2015; Odert et al. 2017). Observations of CMEs themselves are necessary to confidently characterize a star's CME behavior.

To constrain the potential impact of CMEs on HWO's biosignature search, it is also necessary to consider the time required for a planetary atmosphere to chemically recover from a CME/SEP impact. Simulations of the response of an Earth-like atmosphere to events of varying energy indicate ozone columns take ~10 yr to return to pre-event levels following an event above the threshold for impact (Tilley et al. 2019).

**These considerations combine to define a critical threshold:** HWO must be capable of detecting events that occur with energies and rates sufficient to impact ozone chemistry. Specifically, HWO must be able to constrain the rate of events above the threshold of influence to fewer than 0.1 $yr^{-1}$ even in the event that no events are detected. This framing ensures that a meaningful constraint is possible given the true rate of a star's CMEs will not be known in advance, and nondetections due to low CME rates are a plausible, perhaps even a desired, outcome.

Observational monitoring campaigns of a single target lasting 10 years with HWO are not realistic. A compromise is possible by accepting an extrapolation from a constraint on the rate of lower-energy, more frequent events to the rate of events exceeding the threshold of influence. We consider an appropriate balance between extrapolating CME rates and minimizing HWO monitoring time to be setting a detection threshold 2 dex below threshold of influence (in event energy), yielding $10^{31}$ erg for G-type stars and $10^{28}$ erg for late M stars. With this shift, 0.1 (influential) events $yr^{-1}$ translates to 0.3 (detectable) events $d^{-1}$. This assumes a CME rate-energy cumulative distribution follows a log-log power law with a slope of -1.5, mimicking the Sun's.

As an aside: an alternative solution enabling longer monitoring could be to develop a complementary mission capable of wide-field monitoring of many stars to accumulate > 10 star-years of monitoring time for stars similar in type and activity level to the targets of HWO's biosignature search. However, this document will remain focused on the simple solution of using HWO itself to constrain host star CME rates.

For events occurring at a rate of > 0.3 $d^{-1}$, 20 d of monitoring is sufficient to detect at least one event with a likelihood of > 0.01 (assuming Poisson statistics). Therefore, 20 d of monitoring is the threshold required for HWO to limit the likelihood of > 1 influential event in the 10 yrs preceding observation to < 0.01. If events are detected, HWO will provide a constraint on their rate, and subsequent modeling can be applied to determine the impact to biosignature interpretation for the planet in question. As a reference point to gauge the tradeoff between the brevity of monitoring time and the extent of the extrapolation required: extending monitoring time to 100 d would reduce the required extrapolation in event energy to 1.5 dex ($10^{31.5}$ erg for G-type stars, $10^{28.5}$ erg for late M-type stars).

Past CME candidates with energies below these thresholds exist, but their validity was not robustly established—a problem that HWO must overcome. For example, Moschou et al. 2019 estimated kinetic energies as low as $10^{30}$ erg among an ensemble of CME candidates they analyzed. For this reason, enabling science requires that HWO be capable of employing at least two separate diagnostics to validate events. Table 1 lays out high-level criteria that would enable HWO to make contributions to stellar CME science ranging from state of the art to breakthrough.

**Table 1:** High-level capability definitions for stellar CME science.

| **Physical Parameter** | **State of the Art** | **Incremental Progress (Enhancing)** | **Substantial Progress (Enabling)** | **Major Progress (Breakthrough)** |
|---|---|---|---|---|
| CME Detection Confidence | 1-2 diagnostics | 1-2 diagnostics $p < 1\%$ | 2 diagnostics $p < 10^{-3}$ | ≥ 2 diagnostics $p < 10^{-4}$ |
| Energy Detection Threshold for Monitoring | $10^{30}$ erg nearby targets | $10^{31}$ erg G<br>$10^{28}$ erg late M<br>nearby targets | $10^{31}$ erg G<br>$10^{28}$ erg late M<br>HWO targets | $10^{30}$ erg G<br>$10^{27}$ erg late M<br>HWO targets |
| Rate constraint for impactful events | $< 1$ d$^{-1}$ | $< 0.1$ d$^{-1}$ | $< 0.1$ yr$^{-1}$ | $< 0.01$ yr$^{-1}$ |

We have neglected impact probability of CMEs in this formulation, implicitly assuming all CMEs will impact the planet of interest. In reality, the probability of impact will be less than 100%. As a result, the likelihood of chemical impacts derived from a CME rate under this implicit assumption can be regarded as an upper limit. In practice, detailed magnetohydrodynamic modeling based on ground-based Zeeman Doppler Imaging observations of a star's magnetic field may be used to estimate CME trajectories, opening angles, and ultimately impact probabilities (e.g., Kay et al. 2019). For stars where rates of influential CMEs exceed 0.1 yr$^{-1}$, this modeling will be an essential precursor to models of atmospheric chemistry that assess biosignature impacts.

Another limitation of the formulation presented here is its association between CME energy and ozone depletion. In particular, the intermediate step in which the shock driven by a CME generates SEPs that then precipitate onto a planetary atmosphere is complex, as has been shown by solar observations and CME-SEP modeling in solar and stellar contexts (Hu et al. 2022; Cheng et al. 2024; Papaioannou et al. 2024). Two CMEs with the same kinetic energy could produce substantially different SEP events if, for example, they propagate at different speeds. Our approach could be improved by implementing a more sophisticated treatment of the CME-SEP relationship and its influence on ozone and other biosignatures.

Our framing in this section, and this document as a whole, represents an opening attempt at defining CME measurement objectives aimed at mitigating the risk posed by CMEs to HWO's life-finding mission. There is much work in the area of quantifying potential biosignature interpretation risk still to be conducted, such as modeling alternative chemical pathways by which CMEs and associated SEPs might impact exoplanetary atmospheric chemistry that differ from Earth's. The huge volume of parameter space where space weather impacts have yet to be explored, and assumptions that have yet to be tested, is all the more reason to establish a footing for CME observations as early as possible within the HWO science portfolio. The potential risks are substantial, so capturing criteria for hardware and operations designs that would enable CME characterization, even before we understand the full breadth and depth of their potential impacts to biosignatures, is key.

## Step 4: Description of Observations

Detect events with energies above the threshold of influence, or limit their rate to < 0.1 $yr^{-1}$, using at least one, ideally two or more, of the following proposed methods 1) coronal dimming, 2) coronal emission Doppler shifts, 3) high contrast imaging, and 4) planetary aurora. For detected events, measure their velocity and mass with one or more methods, yielding energy estimates with accuracy < 0.5 dex.

We choose this energy accuracy threshold to keep energy uncertainty small relative to the 2 dex energy extrapolations required to infer (or limit) the rate of influential events. However, we note that uncertainties on the parameters of solar CMEs are typically considered to be a factor of 2 (0.3 dex). Future work could explore trading increased monitoring for lower (e.g, 1 dex) energy accuracy.

**Observation details**

1) **Coronal dimming.** Coronal dimming is a decrease in the light emitted by a stellar corona as a result of the dark, rarefied "hole" left behind by a CME. Although solar coronal dimming has been primarily observed in extreme UV and soft X-ray lines and bands, HWO will be capable of observing coronal dimming in far ultraviolet coronal emission lines provided these criteria are met:
    a) Bandpass: Encompass lines near the ambient coronal temperature such as the Fe XII lines at 1242 and 1349 Å ($10^{6.15}$ K) and the Fe XXI 1354 Å ($10^{7.04}$ K) line. Including Fe XIX 1118 Å ($10^{6.89}$ K) may enhance sensitivity

to CMEs, particularly for active stars. Similarly, Fe XVIII 974 Å ($10^{6.81}$ K) and O VI 1032 Å ($10^{5.45}$ K) would provide coverage at more temperatures.

b) Resolution: Lines must be separable. The most stringent case is Fe XXI's natural blend with a C I line of the same wavelength requiring < 0.05 Å resolution (R > 27,000) to deblend. The next most stringent case is the Fe XII and N V near-blend at 1242 Å, requiring < 0.2 Å resolution (R > 6000) to separate.

c) Sensitivity: 0.06% / √hr for young stars, 0.6% / √hr for mature stars (see justification below). This assumes integrations of > 9 h, yielding 3σ detectability for a threshold event. Fluctuations in stellar coronal emission are likely to impose a noise floor of 0.1-1% if not well-modeled. The development of techniques to tease out minor dimming signals from correlated astrophysical noise through advanced activity-variability models, as with exoplanet radial velocity and transit signals, will be essential to enabling these measurements. These precision constraints, combined with estimates of stellar coronal line fluxes for likely target systems, will translate to sensitivity constraints.

d) Time cadence: A cadence of < 10 min is needed to resolve the initial dimming, constraining CME acceleration and speed. Combined with mass estimated from dimming depth, this yields a kinetic energy estimate.

e) Monitoring: A minimum of 3 h is needed to observe the development of solar-like dimmings. However, stares of > 20 h are necessary to observe the full life-cycle of an event, providing out-of-event context.

2) **Coronal emission Doppler shifts** As coronal material is ejected, its emission is blueshifted prior to expansion and quenching. Blueshifts are more pronounced in hotter coronal lines. Characteristic speeds are 100 – 1000 km $s^{-1}$. Speeds may be directly estimated and masses inferred from the amount of emission provided an estimate of the pre-ejection density is available (such as from the Fe XII 1242,1349 Å pair).

a) Bandpass: Same as coronal dimming.

b) Resolution: To clearly distinguish weak emission components shifted as little as 100 km s-1 from the line centroid will require ~30 km s-1 resolution, equating to R > 10,000.

c) Sensitivity: 0.01% / √hr for young stars, 0.1% / √hr for mature stars. These limits are lower than coronal dimming because Doppler shifted emission will rapidly quench as the CME expands (producing a subsequent dimming). Detections at 3σ must be possible with 10 min integrations.

d) Time cadence. A cadence of < 10 min is needed to capture emission before it is quenched.

e) Monitoring: Individual stares lasting > 1 h are needed to provide out-of-event context.

3) **High contrast imaging** Initial estimates suggest stellar coronagraphy with HWO may be more feasible than previously thought for the detection of CMEs. This technique is appealingly direct. However, more work is needed to establish sensitivity requirements. Zodiacal light may prove prohibitive to the detection of CMEs with high contrast imaging, though strong winds from active stars might rapidly clear out zodiacal dust through enhanced drag. Regardless, it is likely that direct imaging of CMEs will only be feasible for CMEs with high masses yielding energies much greater than the threshold of influence. Nonetheless, direct detection of CMEs would be breakthrough science in its own right, and, further, it would enable the validation of the other techniques listed here for stars with much higher activity levels and potentially lower masses than the Sun, enhancing confidence in their application to targets and CMEs where direct CME imaging is not possible.
   a) Bandpass: Broadband optical to observe Thomson scattered light. (Future work might explore UV and optical iron emission lines that could offer higher contrast at the cost of lower fluxes).
   b) Resolution:
      i) Spectral: No requirement.
      ii) Spatial: Roughly ⅓ of the inner working angle to discern a CME from the background stellar wind at the inner limit of the imager.
   c) Sensitivity: More work needed.
   d) Contrast: $10^{-12}$ for sensitivity to Sun-like events, $10^{-10}$ for sensitivity to masses on the order of $10^{19}$ g as have been reported for stellar CME candidates/prominences (e.g. summarized in Moschou et al. 2019, reported in Namekata et al. 2024).
   e) Time cadence: $< d\,\Theta / 3\,v$ where $d$ is distance to star, $\Theta$ is the angular size of the field of view, and $v$ is velocity of the event. This enables at least three frames of the event to establish a speed, provided the inner working angle is small relative to the field of view.
   f) Monitoring: $> d\,\theta / v$ where $d$ is distance to star, $\theta$ is inner working angle, and $v$ is velocity of the event to ensure an event detected via dimming or doppler shift has sufficient time to emerge into the field of view. To further ensure a velocity measurement is possible, an amount 3× the time cadence should be added as well.

4) **Planetary aurorae** The impact of a CME, as well as the SEP event it may launch ahead of it, may incite auroral (or auroral-like) emission in a planetary atmosphere. While estimates of kinetic energy based on auroral emission would be highly model dependent, an aurora following a coronal dimming or Doppler shift signal would be a clear confirmation that a CME occurred. Further, because aurorae result from high energy particles impacting atmospheric molecules, they could provide direct insight into chemical effects on planetary atmospheres.

a) Bandpass: Far UV, especially Lyman-α (1215.67 Å) and/or optical.
b) Spectral resolution: Ability to separate typical emission lines (R > 10,000).
c) Mode: Integral field unit to obtain simultaneous spectroscopy of star and planet during host star monitoring without a-priori knowledge of the planetary position (i.e., enabling a “blind search” for aurorae).
d) Sensitivity: More work needed.
e) Monitoring: > *a / v* where *a* is the planetary semi-major axis and *v* is the velocity of the event — ~10–100 h solar CMEs impacting Earth.

**Additional Details Regarding Estimates**

For coronal dimming and Doppler shifted coronal emission, the driving factor for required sensitives is the ability to detect events above a threshold mass. We have estimated the masses of stellar corona using models of their temperature–density structure that are informed by UV data, yielding values of

- G star, mature: $4 \times 10^{17}$ g ((Fontenla et al. 2016), based on Fig. 1)
- G star, young: $4 \times 10^{18}$ g (assumes an order of magnitude jump, as with M stars below)
- early M star, mature: $10^{16}$ g (Peacock et al. 2020, based on Fig. 8)
- early M star, young: $10^{17}$ g (Peacock et al. 2020, based on Fig. 8)

It is important to note that this is a key area of uncertainty. A large range of basal column densities/basal pressures at the base of stellar coronae can match observed activity-generated emission (Peacock et al. 2020, Fig. 3). More accurate methods of estimating coronal masses and thus dimming depth signals should be explored to improve these estimates. Improved estimates is an anticipated component of the funded ADSPS23-0053 research program. Although we used Trappist-1 as an extreme example above, we have not included late M stars like Trappist-1 here because they are not likely to be targets of HWO’s direct-imaging biosignature search.

Combining these estimates with thresholds of influence based on the habitable zone distance for these stellar types flows to the fraction of coronal luminosity emitted by the CME. This emission is assumed to initially show up as a Doppler shifted emission component in coronal lines and then disappear, resulting in a dimming event of the same depth. Converting a threshold energy to mass requires an assumption of CME velocity. Higher velocities will correspond to lower masses for a fixed energy, amounting to weaker fractional emission. We assume a velocity of 1000 km $s^{-1}$, near the upper end of solar CME velocities (Gopalswamy et al. 2009). Younger, more active stars may be subject to competing effects of more energetic events (increasing CME speeds) and stronger overlying fields (slowing CME speeds). Therefore, we adopt the same velocity for both mature and young cases.

**Table 2:** SNR requirements for coronal dimming.

| Type | Age Class | Inner-HZ Distance (au) | Influential CME Energy* (erg) | Detection CME Energy* (erg) | Min. Velocity (km $s^{-1}$) | CME Threshold Mass (g) | Total Coronal Mass (g) | CME Emission Fraction | SNR / √hr Req'd for 3σ Detection |
|---|---|---|---|---|---|---|---|---|---|
| G2 | mature | 1 | $10^{33}$ | $10^{31}$ | 1000 | $10^{15.3}$ | $10^{17.5}$ | 0.6% | 500 |
| G2 | young | 1 | $10^{33}$ | $10^{31}$ | 1000 | $10^{15.3}$ | $10^{18.5}$ | 0.06% | 5000 |
| M1 | mature | 0.2 | $10^{31.5}$ | $10^{29.5}$ | 1000 | $10^{13.8}$ | $10^{16}$ | 0.6% | 500 |
| M1 | young | 0.2 | $10^{31.5}$ | $10^{29.5}$ | 1000 | $10^{13.8}$ | $10^{17}$ | 0.06% | 5000 |

*Influential CME energy is the energy above which a CME is expected to have a substantial influence on ozone chemistry. The Detection CME energy is the corresponding detection threshold to place a meaningful constraint on the rate of influential CMEs. See Step 3 for explanation and justification.

To provide a more tangible constraint on requirements for HWO, we use coronal emission line estimates for proxy targets adjusted to 10 pc to derive sensitivity requirements in units of flux (Table 2). We scaled upward and downward based on Rossby number, Ro, an independent variable that predicts X-ray emission (Wright et al. 2018) for values representative of young, saturated-activity stars (Ro < 0.1) and mature stars (Ro > 1), based on the approximate Rossby numbers of the proxies (Ro = 0.2 for EK Dra, Ro = 0.3 for $\chi^1$ Ori,  Ro < 0.1 for AU Mic, Ro = 2 for α Cen A). Fe XVIII and Fe XIX measurements are from Redfield et al. (2003), O VI from Redfield et al. (2002), and Fe XII and Fe XXI from Ayres et al. (2003). For future work, it might be possible to improve these estimates by locating stellar proxies with measured emission covering all ages or by considering emission temperature when adjusting by Rossby number (see, e.g., Loyd et al. 2021; Pineda et al. 2021; Wright et al. 2018).

**Table 3:** Typical fluxes of coronal emission lines in log(erg $s^{-1}$ $cm^{-2}$) units.

| Type | Age Class | Proxy | Fe XVIII 974 Å @ 10 pc | Fe XIX 1118 Å @ 10 pc | Proxy | O VI 1032 Å @ 10 pc | Proxy | Fe XII 1349 Å @ 10 pc | Fe XXI 1354 Å @ 10 pc |
|---|---|---|---|---|---|---|---|---|---|
| G2 | mature | 0.05 × EK Dra | -15.2 | -15.5 | α Cen A | -13.7 | 0.1 × $\chi^1$ Ori | -15.4 | -15.6 |
| G2 | young | 5 × EK Dra | -13.2 | -13.5 | 100 × α Cen A | -11.7 | 10 × $\chi^1$ Ori | -13.4 | -13.6 |
| M1 | mature | 0.01 × AU Mic | -16.2 | -16.2 | 0.01 × AU Mic | -14.7 | 0.01 × AU Mic | -16.9 | -16.2 |
| M1 | young | AU Mic | -14.2 | -14.2 | AU Mic | -12.7 | AU Mic | -14.9 | -14.2 |

Combining the SNR requirements from Table 2 and the typical line fluxes from Table 3 leads to the FUV sensitivity requirements shown in Table 4.

**Table 4:** HWO sensitivity requirements (1σ noise floor in log(erg $s^{-1}$ $cm^{-2}$) units).

| Type | Age Class | Fe XVIII 974 Å @ 10 pc | Fe XIX 1118 Å @ 10 pc | O VI 1032 Å @ 10 pc | Fe XII 1242,1349 Å @ 10 pc | Fe XXI 1354 Å @ 10 pc |
|---|---|---|---|---|---|---|
| G2 | mature | -17.9 | -18.2 | -16.4 | -18.1 | -18.3 |
| G2 | young | -16.9 | -17.2 | -15.4 | -17.1 | -17.3 |
| M1 | mature | -18.9 | -18.9 | -17.4 | -19.6 | -18.9 |
| M1 | young | -17.9 | -17.9 | -16.4 | -18.6 | -17.9 |

Table 3 can be interpreted as a trade space, showing how sensitivity and bandpass will influence the range of stellar types and ages for which HWO could constrain the rate of influential CMEs. Coronal dimming appears most clearly in emission from plasma near the peak temperature of a stars' coronal emission measure distribution. This temperature progresses from hot to cool as stars age and magnetic activity declines (Johnstone & Güdel 2015). We have highlighted in green the emission lines with formation temperatures best suited for measuring coronal dimming for each case in Table 4 accordingly.

Mature M dwarfs represent the most constraining case for observing coronal dimming. O VI offers a bright option for a limited sample of stars with particularly weak activity. If focusing only on old G stars, a sensitivity limit of ~$10^{-18}$ erg $s^{-1}$ $cm^{-2}$ $Å^{-1}$ (assuming a 1 Å, ~300 km $s^{-1}$ line width) at 1350 Å may suffice and will also make measurements possible for young G and M stars. When considering the Fe XII sensitivity values, note that it would be greatly beneficial to include both the 1242 and 1349 Å lines in the bandpass, as they represent a density-sensitive pair, greatly reducing uncertainty in CME masses derived from coronal dimming.

To adapt these estimates to measuring doppler shifts in emission, we reduce required precision by a factor of 6 to enable measurements in at a 10 minute, rather than 1 hour, time resolution.

**Justification for HWO**

Other observatories aside from HWO could yield constraints on stellar CMEs. However, this would make HWO dependent on these observatories to obtain the full context it needs to interpret signs of life or its absence, adding mission risk. Moreover, other science cases are likely to drive the instrument criteria needed to enable HWO to observe stellar CMEs, and the monitoring intended for target systems will provide ample opportunity to do so. It is even possible that only HWO will have the capability of observing two or more CME diagnostics from the same platform, perhaps even tracking the life cycle of a stellar CME, producing Doppler shifted emission quickly followed by dimming as it is launched, followed hours to days later by auroral emission on orbiting planets as the CME propagates outward. Therefore, we deem it highly worthwhile to consider CME detention strategies as part of HWO's science portfolio.

**Design criteria**

In Table 5, we give design criteria. We consider detecting CMEs for Sun-like stars with the dimming method alone to be "enhancing", while the sensitivity required to employ dimming and Doppler shifts with a wider array of lines to be "enabling", and opening up more techniques (aurorae, direct imaging) with sensitivity to old early M stars would be "breakthrough."

**Table 5:** Design criteria.

| **Observation Requirement** | **State of the Art** | **Incremental Progress (Enhancing)** | **Substantial Progress (Enabling)** | **Major Progress (Breakthrough)** |
|---|---|---|---|---|
| Type | Spectroscopy | Spectroscopy | Spectroscopy | IFU Spectroscopy + High Contrast Imaging |
| Wavelength Range | 1240 - 1400 Å | 1240 - 1400 Å | 970 - 1400 Å | 970 - 1400 Å + optical |
| Sensitivity (erg $s^{-1}$ $cm^{-2}$ $Å^{-1}$ noise floor) | $10^{-17}$ | $10^{-18}$ | $10^{-19}$ | $10^{-20.5}$ |
| Time Cadence | < 10 min | — | — | — |
| Unbroken Stares | 5 - 10 h | > 10 h | > 100 h | > 1000 h |
| Cumulative Star-Time* | 10 $d^{-1}$ | > 10 d | > 20 d | > 100 d |
| Contrast | | | | < $10^{-10}$ |
| Supporting Work** | S2 solar | S2 nonsolar | S1 & S2 nonsolar | S1 & S2 nonsolar |
| Additional Sensitivity (More Work Needed) | | | | CME Thomson scattering or aurorae detectable |

*For stars of similar type and activity to the biosignature target. Achieving > 100 d star time might be possible with an IFS with a field of view that enables simultaneous spectral monitoring of more than one star.
** S1) Advanced models of stellar quiescent variability. S2) Quantification of false/true negatives/positives (a diagnostic matrix) in either or both solar context (with observations) and stellar context (with modeling or via cross-validation).

# References

Airapetian, V. S., Barnes, R., Cohen, O., et al. 2020, International Journal of Astrobiology, 19, 136
Airapetian, V. S., Glocer, A., Gronoff, G., Hébrard, E., & Danchi, W. 2016, Nature Geoscience, 9, 452
Ayres, T. R., Brown, A., Harper, G. M., et al. 2003, The Astrophysical Journal, 583 (IOP), 963
Calisto, M., Usoskin, I., & Rozanov, E. 2013, Environmental Research Letters, 8 (IOP), 045010
Chen, H., De Luca, P., Hochman, A., & Komacek, T. D. 2025 (arXiv), https://ui.adsabs.harvard.edu/abs/2025arXiv250503723C
Chen, H., Zhan, Z., Youngblood, A., et al. 2021, Nat Astron, 5 (Nature Publishing Group), 298
Cheng, L., Zhang, M., Kwon, R. Y., & Lario, D. 2024, arXiv e-prints, https://ui.adsabs.harvard.edu/abs/2024arXiv241104095C
Chi, Y., Shen, C., Wang, Y., et al. 2016, Sol Phys, 291, 2419
Dong, C., Lingam, M., Ma, Y., & Cohen, O. 2017, The Astrophysical Journal, 837 (IOP), L26
Drake, J. J., Cohen, O., Garraffo, C., & Kashyap, V. 2015, Proc IAU, 11, 196
Edberg, N. J. T., Nilsson, H., Futaana, Y., et al. 2011, J Geophys Res, 116, n/a
Edberg, N. J. T., Nilsson, H., Williams, A. O., et al. 2010, Geophysical Research Letters, 37 (Wiley), L03107
Espinoza, N., & Perrin, M. D. 2025 (arXiv), https://ui.adsabs.harvard.edu/abs/2025arXiv250520520E
Fontenla, J. M., Linsky, J. L., Garrison, J., et al. 2016, The Astrophysical Journal, 830 (IOP), 154
France, K., Duvvuri, G., Egan, H., et al. 2020, The Astronomical Journal, 160 (IOP), 237
Gao, P., Wakeford, H. R., Moran, S. E., & Parmentier, V. 2021, Journal of Geophysical Research (Planets), 126 (Wiley), e06655
Gopalswamy, N., Yashiro, S., Michalek, G., et al. 2009, Earth Moon and Planets, 104 (Springer), 295
Howard, W. S., Kowalski, A. F., Flagg, L., et al. 2023, The Astrophysical Journal, 959 (IOP), 64
Hu, J., Airapetian, V. S., Li, G., Zank, G., & Jin, M. 2022, Science Advances, 8 (American Association for the Advancement of Science), eabi9743
Johnstone, C. P., & Güdel, M. 2015, Astronomy and Astrophysics, 578, A129
Kay, C., Airapetian, V. S., Lüftinger, T., & Kochukhov, O. 2019, ApJL, 886 (The American Astronomical Society), L37
Khodachenko, M. L., Ribas, I., Lammer, H., et al. 2007, Astrobiology, 7, 167
Kobayashi, K., Airapetian, V. S., Udo, T., et al. 2026 (arXiv), https://ui.adsabs.harvard.edu/abs/2026arXiv260318206K
Kobayashi, K., Ise, J., Aoki, R., et al. 2023, Life, 13, 1103
Lammer, H., Lichtenegger, H. I. M., Kulikov, Y. N., et al. 2007, Astrobiology, 7, 185
Lammer, H., Zerkle, A. L., Gebauer, S., et al. 2018, Astronomy and Astrophysics Review, 26 (Springer), 2

Lingam, M., Dong, C., Fang, X., Jakosky, B. M., & Loeb, A. 2018, The Astrophysical Journal, 853, 10
Loyd, R. O. P., Shkolnik, E. L., Schneider, A. C., et al. 2021, The Astrophysical Journal, 907, 91
Luhmann, J. G., Kasprzak, W. T., & Russell, C. T. 2007, J Geophys Res, 112, E04S10
Miyake, F., Nagaya, K., Masuda, K., & Nakamura, T. 2012, Nature, 486, 240
Moran, S. E., Hörst, S. M., Vuitton, V., et al. 2020, The Planetary Science Journal, 1 (IOP), 17
Moschou, S.-P., Drake, J. J., Cohen, O., et al. 2019, The Astrophysical Journal, 877, 105
Namekata, K., Airapetian, V. S., Petit, P., et al. 2024, The Astrophysical Journal, 961, 23
Namekata, K., Maehara, H., Notsu, Y., et al. 2025, The Astrophysical Journal, 993 (IOP), 80
Odert, P., Leitzinger, M., Hanslmeier, A., & Lammer, H. 2017, Monthly Notices of the Royal Astronomical Society, 472, 876
Papaioannou, A., Herbst, K., Ramm, T., et al. 2023, A&A, 671 (EDP Sciences), A66
Papaioannou, A., Herbst, K., Ramm, T., Lario, D., & Veronig, A. M. 2024, Astronomy and Astrophysics, 690 (EDP), A60
Peacock, S., Barman, T., Shkolnik, E. L., et al. 2020, The Astrophysical Journal, 895, 5
Pineda, J. S., Youngblood, A., & France, K. 2021, The Astrophysical Journal, 911, 111
Raeside, S. R., Rodgers-Lee, D., & Rimmer, P. B. 2025 (arXiv), http://arxiv.org/abs/2504.02596
Ramstad, R., & Barabash, S. 2021, Space Sci Rev, 217, 36
Redfield, S., Ayres, T. R., Linsky, J. L., et al. 2003, The Astrophysical Journal, 585 (IOP), 993
Redfield, S., Linsky, J. L., Ake, T. B., et al. 2002, The Astrophysical Journal, 581 (IOP), 626
Segura, A., Walkowicz, L. M., Meadows, V., Kasting, J., & Hawley, S. 2010, Astrobiology, 10, 751
Tabataba-Vakili, F., Grenfell, J. L., Grießmeier, J.-M., & Rauer, H. 2016, A&A, 585 (EDP Sciences), A96
Tilley, M. A., Segura, A., Meadows, V., Hawley, S., & Davenport, J. 2019, Astrobiology, 19, 64
Tomasko, M. G., Archinal, B., Becker, T., et al. 2005, Nature, 438, 765
Venot, O., Rocchetto, M., Carl, S., Roshni Hashim, A., & Decin, L. 2016, The Astrophysical Journal, 830 (IOP), 77
Wright, N. J., Newton, E. R., Williams, P. K. G., Drake, J. J., & Yadav, R. K. 2018, Monthly Notices of the Royal Astronomical Society, 479, 2351
Yung, Y. L., Allen, M., & Pinto, J. P. 1984, The Astrophysical Journal Supplement Series, 55 (IOP), 465